\documentclass[11pt]{article} 
\usepackage[utf8]{inputenc}

\usepackage{amsfonts}
\usepackage{amsmath}
\usepackage{amssymb}
\usepackage{amsthm}
\usepackage{caption}
\usepackage{color}
    \definecolor{darkgreen}{rgb}{0,0.5,0}
    \definecolor{darkblue}{rgb}{0,0,0.6}
    \definecolor{purple}{rgb}{0.4,.2,0.7}
\usepackage[margin = 2.5cm]{geometry}
\usepackage{graphicx}
\usepackage[hyperfootnotes = false, colorlinks = true, linkcolor = darkblue, citecolor = purple]{hyperref}
\usepackage{subcaption}
 
\newcommand{\be}{\begin{equation}}
\newcommand{\ee}{\end{equation}}
\newcommand{\bea}{\begin{eqnarray}}
\newcommand{\eea}{\end{eqnarray}}
\def\la{\label}
\def\nref#1{(\ref{#1})}
\def\half{{1 \over 2 }}

\begin{document}

\thispagestyle{empty}
\begin{center}
    ~\vspace{5mm}
    
    {\LARGE \bf {Comments on magnetic black holes \\}} 
    
    \vspace{0.5in}
    
    {\bf   Juan Maldacena}

    \vspace{0.5in}

    Institute for Advanced Study,  Princeton, NJ 08540, USA \vskip1em
     
    \vspace{0.5in}
    

\end{center}

\vspace{0.5in}

\begin{abstract}
 We discuss aspects of magnetically charged black holes in the Standard Model. For a range of charges,  we argue that the electroweak symmetry is restored in the near horizon region. The extent of this phase can be macroscopic. If $Q$ is the integer magnetic charge,  the fermions lead to order $Q$ massless two dimensional fermions moving along the magnetic field lines. These greatly enhance Hawking radiation effects.

\end{abstract}

\vspace{1in}

\pagebreak

\setcounter{tocdepth}{3}


\section{Introduction} 
 
 We discuss some properties of magnetic black holes in the Standard Model.  These are solutions of the laws of physics as we know them, though they  do not seem easy to produce. Nevertheless,  they are worth exploring,  since they have interesting features.   
  
     A magnetic black hole is a black hole with a magnetic charge. It is a solution of the Standard Model coupled to gravity. More precisely, we need to assume that the $U(1)$ gauge group is really $U(1)$ and not $R$.  
  It can be trusted as a solution as long as the integer magnetic charge is very large, $Q \gg 1$. So it can be viewed a type of magnetic monopole that does not require any new physics. 
  
   We will discuss extremal and near extremal black holes. Extremal magnetic black holes are more stable than their electric counterparts, since unit charge magnetic monopoles are much heavier than electrically charged particles, and are thus harder to pair create. In addition, an electrically charged black hole can be neutralized in conductive medium, while a magnetically charged one cannot be neutralized with ordinary matter.   
  
  The magnetic fields near the horizon can be very large. So these black holes naturally provide a physical setup with very large  magnetic fields. 
  These large fields can restore the   electroweak symmetry  in the near horizon limit. In fact, for a range of magnetic charges,  we have a  black hole carrying only weak hypercharge magnetic field that is surrounded by a larger a ``electroweak corona", 
  where the Higgs field changes from zero to its usual non-zero value outside, see figure \ref{BHstructure}. The region with restored electroweak symmetry can be much larger than the electroweak scale. For example, for a magnetic black hole of a mass of about the earth, it can be a few millimeters.

 The standard model fermions in the presence of these magnetic fields develop Landau levels, the lowest with zero energy. This lowest level has a degeneracy of order $Q$. This leads to order $Q$ light modes that can go in and out of the black hole. For this reason, phenomena like Hawking radiation are enhanced by a factor of the charge of the black hole. 
 This accelerates the evaporation of such black holes. For example, a near extremal magnetic black hole with the mass of mountain ($10^{12}$ kg) would decay  to extremality with a time scale of order milliseconds, while a Schwarzschild black hole of the same mass would take a  time of the order of the age of the universe.  
 
 Because black holes do not preserve baryon number, these black holes can catalyze proton decay. Protons that fall into the black holes could be returned as positrons. The $Q$ effectively two dimensional massless modes provide a large enhancement to this process, both because they are massless and  because of they are many. 
 
 In fact,  
 the physics of these black holes has many properties in common with magnetic monopoles. What we described in the previous paragraph is similar to the Callan-Rubakov effect for monopoles \cite{Rubakov:1981rg,Rubakov:1982fp,Callan:1982ah,Callan:1982au,Callan:1982ac}. 

 In the remainder of this paper we discuss some of these properties in more detail, but leave many unanswered questions for the future.   This paper expands on some comments made in \cite{Maldacena:2018gjk}, whose main focus was different.

 \section{Classical Magnetic black holes }

 The starting point is the standard extremal charged black hole solution (or Reissner Nordstr\"om solution) 
 \bea
 ds^2 &=& - f dt^2 + { dr^2 \over f } + r^2 (d\theta^2 + \sin^2 \theta d\varphi^2 ) ~,~~~~~f = \left(1- {r_e \over r } \right)^2 ~,~~~~~~
 \\
 F &=&  d A =  { Q \over 2 }  \sin \theta d\theta d \varphi ~,~~~~|F| = e |B| = \half { Q \over r^2 } ~,~~~~~~~  \vec B  = { Q \vec r \over 2 e  r^3 } \la{Fval}
 \eea 
 where $Q$ is the (integer) magnetic charge in quantized units, 
 $\int_{S^2} F = 2 \pi Q$, and we are imagining that 
 $A$ couples to the unit electric charge via $e^{ i \int A}$.  The mass and size  of the black hole are 
 \be \la{MQr}
 M = { r_e \over l_p^2}  ~,~~~~~~~~~~~~~ r_e \equiv   {Q \sqrt{\pi}\, l_p \over e } ~,~~~~~~~~{\rm with}~~~~~ l_p \equiv \sqrt{G_N}
 \ee
 where $e$ is the electric coupling constant. The conventional magnetic field is given by $B_i = { 1 \over e } \epsilon_{ijk}F_{jk}$. By $|F|$ we denote the proper size of the magnetic field. 
 
 This solution contains two important scales, one is  the extremal radius $r_e$ defined above. The other is set by the square root of the proper size of the magnetic field, $\sqrt{|F|} =\sqrt{ e|B|}$. This second scale is  important when we have charged matter fields as it sets the energy scale of the Landau levels. 
  
  If we have a magnetic field $F_{12}$ in the 12 plane,  the lowest energy Landau level for a particle of spin $s$ leads to the following energies 
  \be \la{LLL}
   E^2 -P_3^2    =
    m^2 + |F| (1  -  2 s ) =  \left\{ 
   \begin{array}{lcl} m^2 +|F|~, & {\rm for } & s =0  
   \cr 	
  m^2+ \, 0~, & {\rm for } & s  =\half   
   \cr 
   m^2- |F|~, &{\rm for }&s = 1  
      \end{array} \right. ~,~~~~~~~~~|F| =|F_{12}|= e|B|
\ee 
 where the spin one result arises when the field is part of a Yang Mills action. All these results are for the case of unit charge. If the field has a different charge we just substitute $F\to q F$, with $q$ the charge in units of the electron charge.  Additional Landau levels are spaced in units of   $2|F|$. Each of these Landau levels has degeneracy $q Q$ where $Q$ is the total flux of $F$. Alternatively we can say that there is one level per unit flux quantum of area. These levels are fairly localized in the transverse space. We can qualitatively picture these modes as traveling along ``wires'' laid down along the magnetic field lines, one wire per flux quantum.  
 
 \subsection{Large magnetic fields in the electroweak theory } 
  
  The proper size of the  magnetic field is highest in the $AdS_2 \times S^2$ near horizon region where it goes as 
  \be \la{MagInt}
  e|B| =  |F|  = { Q \over 2 r_e^2 } = { e^2 \over 2 \pi l_p^2 Q } 
   \ee 
  Note  that this maximum value of the  magnetic field becomes smaller for larger charges. Starting with a very large charge, we will have a relatively low magnetic field. As we reduce the charges the magnetic field becomes stronger. This 
  has various impacts on the physics. 
   
   First we are going to focus on its impact on the electroweak vacuum. When the magnetic field is much larger than the electroweak scale, the electroweak symmetry is restored and only the hypercharge component of the magnetic field survives. This was discussed in a  series of papers by 
   Ambj{\o}rn and Olesen 
   \cite{Ambjorn:1992ca,Ambjorn:1989sz,Ambjorn:1989bd}, see also \cite{Tornkvist:1992kh,Chernodub:2012fi} for more recent discussions. We will review here the basic physics and will discuss how it applies to the solution around a magnetic black hole. These magnetic black holes provide a  ``natural" setup for very large magnetic fields\footnote{Another ``natural'' situation are  superconducting cosmic strings \cite{Witten:1984eb}, as already pointed out in \cite{Ambjorn:1989sz}. }.    
      
   First we note that when the field exceedes the value 
   \be 
    |F|_W \equiv m_W^2 
    \ee 
    the $W$ boson becomes unstable  condenses. If the field is only slightly larger than this value, the $W$ boson will only slightly condense, stopped by higher order terms in the action. As the field increases, this condensation will eventually remove the $SU(2)$ component of the magnetic field. Recall that an ordinary magnetic field is the superposition of two equal magnetic fields $F= F^Y = F^3$, where $F^Y$ is the weak hypercharge magnetic field and $F^3$ is the $U(1) \subset SU(2)$ one. For sufficiently large values of the magnetic field we expect that only the hypercharge component survives. 
     Since the Higgs boson has hypercharge $q_Y=\half$,   it has a large energy   in the presence of a purely hypercharge magnetic field  \nref{LLL}. This   drives the Higgs vacuum expectation value to zero for sufficiently large fields. If we start from the vacuum with zero Higgs vacuum expectation value, we see from \nref{LLL} that 
     it will be stable for a  magnetic field   larger than 
     \be 
      \half |F|_H = \mu^2 ~,~~~~~~|F|_H \equiv m_H^2 
      \ee
      where the factor of $\half$ arises due to the weak hypercharge of the Higgs field, $ \mu^2$ is the quadratic term in the Mexican hat potential, and we used that $\mu^2  = \half m_H^2$ (at tree level). If the field is only slightly smaller than $|F|_H$ we expect only a small amount of condensation of the Higgs field, since the condensation is stopped by the quartic potential. 
      In nature, $m_H > m_W$ so that $|F|_H>|F|_W$.  
      As explained in \cite{Ambjorn:1992ca,Ambjorn:1989sz,Ambjorn:1989bd}, this then means that, as we increase the magnetic field, the $W$ bosons start condensing and the Higgs expectation value starts decreasing so that, by the time that we reach $|F|_H$, the Higgs expectation is zero. In fact, one can say that \cite{Ambjorn:1992ca}
      \be \la{Window}
       \langle |h|^2 \rangle \sim v^2 { ( m_H^2 - |F|) \over m_H^2 - m_W^2 }  ~,~~~~~~~~~{\rm for }~~~~~~ m_W^2 < |F| < m_H^2 
       \ee 
       where $v$ is the usual Higgs vacuum expectation value. The average in the left hand side includes a spatial average over the transverse directions. In fact, 
       both the $W$ and Higgs condensates break the translation symmetry in the transverse space.  For small Higgs vacuum expectation values this can also be understood in terms of the monopole harmonics, see e.g. \cite{Shnir:2011zz}\footnote{More mathematically,   for small Higgs values, each complex component of the Higgs field is   a section of a holomorphic line bundle   which should have $Q/2$ zeros.}. One can view this as the formation of vortices.   
       
       In other words,   the solution spontaneously breaks translation  symmetry in the transverse space (or spherical symmetry for our application\footnote{For the particular case of $Q=2$,  a spherically symmetric solution was found in \cite{Cho:1996qd}. It contains a singular,  $Q=2$ magnetic hypercharge Dirac monopole surrounded by an $SU(2)$ monopole.}) when the fields are in the window \nref{Window}. 
        In \cite{Ambjorn:1989sz}, these solutions were found in detail for the particular case that $m_Z= m_H$. For larger values of $m_H$, as we have in nature, we expect a similar picture.  
       
       The black hole charge for which we start having a change in the electroweak vacuum  is such that the size of the magnetic field in the near horizon region is about the mass squared of the $W$ boson. 
   Its charge, mass and radius are given by 
    \be \la{EWtrans}
 Q_{ew} =   { e^2 \over 2 \pi  l_p^2 m_W^2 }  \sim  3 \, 10^{32}  ~,~~~~~~  M(Q_{ew}) \sim 4 \,10^{25}\,  {\rm kg} ~,~~~~~~~~ r_e(Q_{ew})  \sim 3 \, {\rm cm}   ~~~~~~~
  \ee     
   When the charge is a bit smaller, so that the field strength is about the Higgs mass squared,   we  have the unbroken phase in the near horizon region. Notice that, for these charges,  the size of the region where the electroweak vacuum is restored has a macroscopic size. 
   
  As we continue to shrink the value of $Q$, $Q\ll Q_{ew}$, then  the precise solution depends on the spectrum of the theory. If we have  the Standard Model and nothing else up to the GUT scale, then  we can continue shrinking the black hole up to relatively small value of $Q$. When the flux becomes comparable to the $X$ and $Y$ boson masses we expect that the black hole will decay into a cloud of magnetic monopoles and become a non-extremal black hole. We will consider charges larger than this lower value in this paper. 
 Solutions in this spirit were discussed by \cite{Lee:1994sk}. Condensation of hair for charged black holes in Anti-de-Sitter space  was extensively studied starting with \cite{Gubser:2000ec}.
     
     Of course, if we had new physics below the GUT scale, it can modify the properties of the solution. Thus, these objects offer a window to very high energy physics. 
     
   \subsection{The electroweak corona} 
  
  When $Q\ll Q_{ew}$ the near horizon region is in the unbroken phase. This phase continues outside until the magnetic field decays to $F \sim m_H^2$. From \nref{Fval} we see that
   this occurs at a distance, $r_h$, given by 
     \be \la{rcdef}
 r_h  = \sqrt{ Q \over 2} { 1 \over m_H}  \gg { 1 \over m_H}  ~,~~~~~{\rm for} ~~~~~~~ Q_{ew} \gg Q\gg 1
 \ee
 where we noted that for large charges this region is much larger than the electroweak scale. 
 At this distance the transition region begins. It ends where the field is equal to $m_W^2$. 
 \be \la{rcdef}
 r_w  = \sqrt{ Q \over 2} { 1 \over m_W}   = r_h { m_H \over m_W} \sim 1.6 \, r_h 
 \ee
These two distances are close because the ratios of masses is close. Note however, that for large $Q$ the thickness of the transition region is much larger than $1/m_H$ by a factor of order $\sqrt{Q}$. It is thicker because the magnetic field is slowly decreasing as we move out. 
 \begin{figure}[t]
    \begin{center}
    \includegraphics[scale=.3]{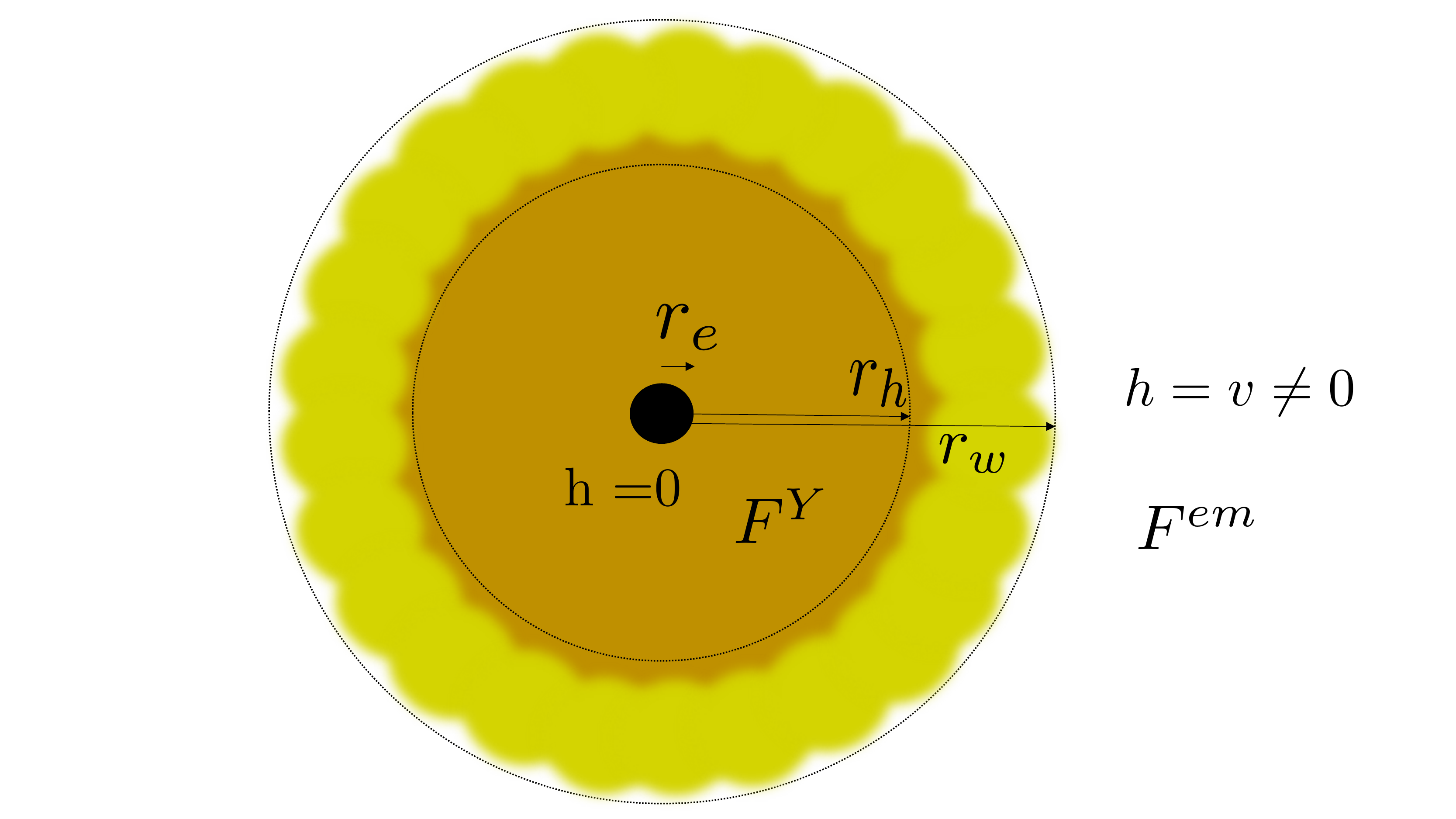}
    \end{center}
    \caption{ Structure of a magnetic black hole for $1\ll Q\ll Q_{ew}$. The black circle represents the near horizon region where the metric is very different from flat space. The brown region, $r< r_h$,  is where the electroweak symmetry is restored and where we have only weak hypercharge magnetic field. The yellow region, $r_h < r < r_w$,  represents the electroweak ``corona''. In this region we have nondiagonal $SU(2)$ gauge fields and a varying higgs field. Further away, $r> r_w$,  we have the usual vacuum with a magnetic field for ordinary electromagnetism, and a  nonzero value for the higgs field.     }
    \label{BHstructure}
\end{figure}

In summary, for $r< r_h$ we have a region with zero Higgs vacuum expectation value and only hypercharge magnetic field. For $r_w < r$ we have a region with the usual Higgs vacuum expectation value and ordinary (electromagnetic) magnetic field. Between $r_h < r < r_w$ we have a transition region that we  call the 
``electroweak corona'' (named in analogy to the solar ``corona'' where magnetic fields play an important role). This is summarized in figure \nref{BHstructure}.  The corona is not spherically symmetric. 

\begin{figure}[t]
    \begin{center}
    \includegraphics[scale=.7]{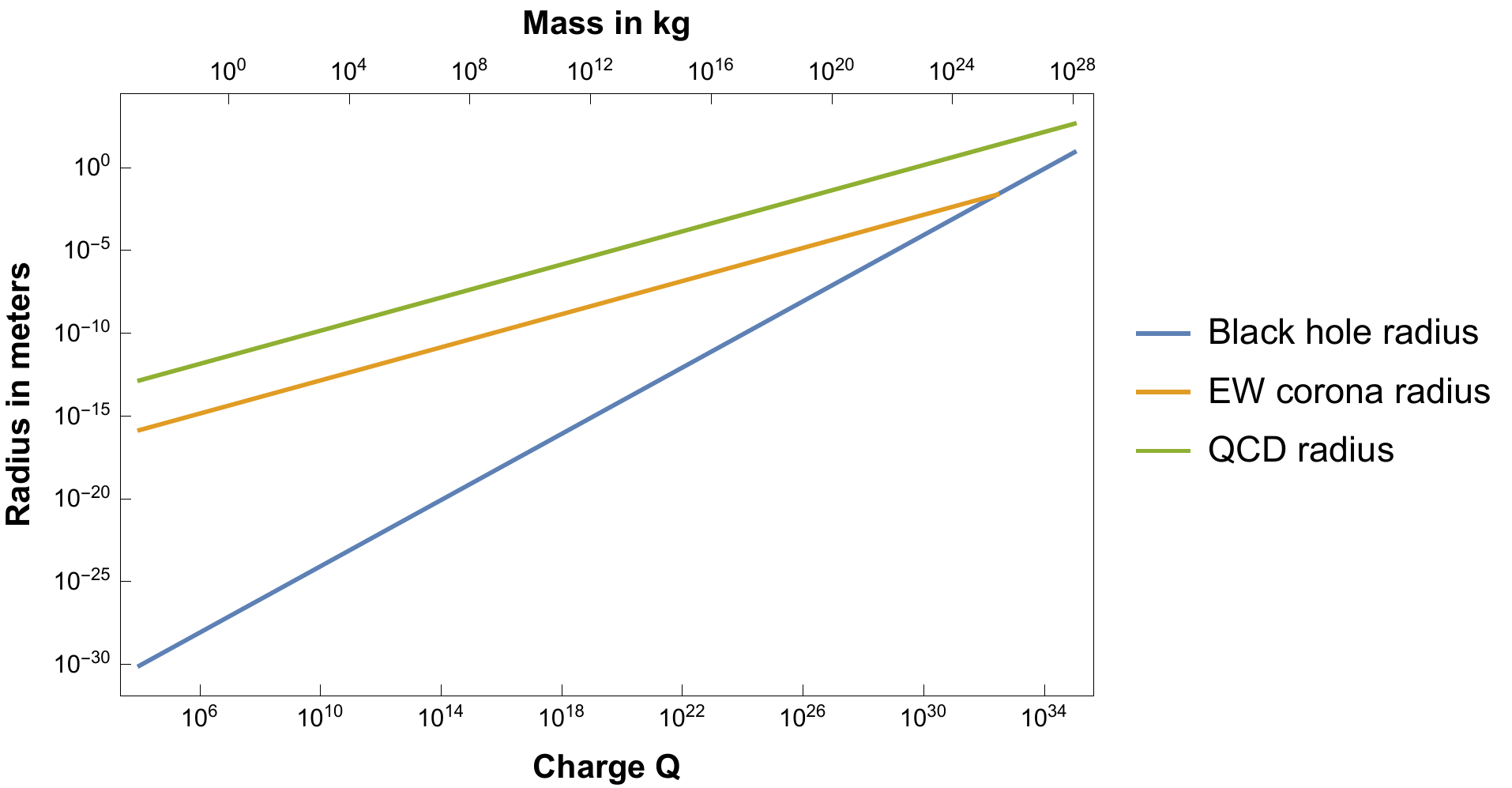}
    \end{center}
    \caption{ We plot the black hole radius $r_e$ and the radius  of the electroweak corona  ($r_c \equiv \sqrt{r_w r_h}$) as a function of the charge. They meet when $r_c \sim  r_e \sim  r_{ew}$ \nref{EWtrans}. We also plotted the  the radius of the region with large deformations to the QCD vacuum.  }
    \label{RadiusPlots}
\end{figure}

For quick reference, in figure \ref{RadiusPlots} we display a plot of the size of the corona $r_c \equiv \sqrt{r_w r_h} $ and also the black hole radius $r_e$ for various charges.

 It is important to note that this electroweak corona will exist for any  localized source of high magnetic charge. The fact that we have a black hole is not important. We get the same ``corona'' for a large $Q$ Dirac monopole for weak hypercharge. In fact, for $Q=2$ an explicit solution was found in \cite{Cho:1996qd}. There is no corona  for the $Q=1$ magnetic monopole because in this case we cannot screen the $SU(2)$ component of the gauge field. Note that for low values of the charges the magnetic field varies over the same scale as its value, while for larger charges it varies more slowly by  a factor of $1/\sqrt{Q}$.

\subsection{Energy of the solution} 

Notice that the original extremal magnetic  black hole with purely electromagnetic fields is a solution for any $Q$.  It has an extremal charge to mass ratio which leads to a zero force condition for equal sign charges. For  $Q< Q_{ew}$ we can condense the $W$ bosons and restore the gauge symmetry. This lowers the energy of the solution.  Therefore these configurations have energies {\it less} than the extremal energy, as expected from the weak gravity conjecture \cite{ArkaniHamed:2006dz}. 
 
 The region near the horizon contributes with a mass as in 
 \nref{MQr} but with $e \to g'$, with $g' = e/\cos \theta_W$ where $\theta_W$ is the  Weinberg angle. Since $g'> e$  the mass from the analog of \nref{MQr} is smaller by a factor of $\cos \theta_W$\footnote{In principle, in formula \nref{MQr} we should use the running coupling $g'$ at the scale $r_e$. This also goes in the direction of making the black hole lighter for smaller masses.  }. 
 We should also add the energy contained in the electroweak corona. 
 This energy can be estimated by computing the contribution due to the fact that the   Higgs is not at the minimum of the potential,
  \be \la{Ecor}
 E_c \propto m_H^4 r_h^3 \propto m_H Q^{3/2}
 \ee 
   Recall that the corona only forms when    $r_w > r_e$. 
 This condition , implies that the corona energy is smaller than the energy gain resulting from the $e \to g'$ replacement in the black hole contribution.  
 
  This implies that two objects with the same  sign magnetic charge  will repel  each other at long distances. \nref{Ecor} suggests that we would minimize the energy by breaking up the black hole into smaller charges. However, this can only happen by tunneling with a suppressed amplitude which we discuss next. 
  
 \subsection{Decay rate} 
 
 These objects are metastable. 
 The magnetic field can decay be creating pairs of magnetic monopoles. Even if magnetic monopoles do not exist, they could decay by emitting smaller magnetically charged black holes. 
 This decay rate can be easily computed in the near horizon region by approximating the monopoles as point particles.  We expect it to be fastest for the lightest object. And we get a rate which is exponentially suppressed as 
 \be \la{Decra}
 \Gamma \propto e^{ - S} , ~~~ S >  \pi Q ( M_{mon} l_p)^2  
  \ee
  where in the last expression we assumed that the mass of the monopole is smaller than the mass of the extremal black hole ($M_{ext}$) with the same charge (see \nref{MQr} for $Q=1$).  For a GUT monopole,  $M_{mon} \sim 10^{17} $Gev,  we find that for $ Q> 10^6 $ the lifetime is larger than the age of the universe. We get a similar   answer if we assume that the unit charge magnetic black hole has a mass near the   weak gravity upper bound \cite{ArkaniHamed:2006dz}. 
  Of course, if the unit charge monopoles are lighter, then one would need to make the charge larger, and the black holes bigger, to ensure long lifetimes.

\section{Matter fields in the black hole background}

First, let us discuss more carefully the issue of charge quantization. 
 For integer magnetic charges, the hypercharge flux is also quantized in terms of integers. However, there are fields of the standard model that carry $1/6$ units of weak hypercharge. This means that we either need to make the hypercharge flux (or original magnetic charge) a multiple of six, or we should include discrete $SU(2)$ or $SU(3)$ fluxes, as is  the case  when the gauge group is really $[U(1) \times SU(2) \times SU(3)]/Z_6$, which is what we get from GUTs. 
  Since we are considering large charges anyway, we will simply assume that they are multiples of six, $Q \in 6 Z$. The other cases can also be treated and we expect similar behavior since the discrete $SU(2)$ and $SU(3)$ fluxes are subdominant for large $Q$. 
  Note, however, that the unit magnetic charge GUT monopole has such fluxes. 
  
  Let us first discuss the fermions in the region where electroweak symmetry is restored. Here all fermions are massless.  
  The magnetic field is very high, and for low energies, the only surviving  modes 
  are the ones  discussed in \nref{LLL} (with $m=0$). In this approximation, these modes are exactly massless but they move only along the radial direction (and the time direction). We can picture them as moving along magnetic field lines. These light modes were already observed for a magnetic monopole 
  \cite{Rubakov:1981rg,Rubakov:1982fp,Callan:1982ah,Callan:1982au,Callan:1982ac}. A new interesting feature for these black holes is that the total number of massless two dimensional modes scales like $Q$.      As we mentioned above, these modes move along magnetic field lines. These magnetic field lines are acting as somewhat similar to superconducting strings, as in \cite{Witten:1984eb}. 
      
      The precise cancellation between orbital   and magnetic dipole energy seems    surprising, but it is explained simply in terms of anomalies. Viewing the background gauge field as non-dynamical, the four dimensional anomaly of a massless Weyl fermion descends to a two dimensional anomaly, which requires massless modes.  
      Each four dimensional Weyl fermion of weak hypercharge $q_Y$ gives rise to 
       \be 
   N = q_Y  Q 
   \ee
   complex two dimensional massless modes. 
  These fermions are   right or left moving  depending on the product of their four dimensional chirality and the sign of their hypercharge. In our four dimensional setup, these $N$ modes form a
  representation of spin $j$, with $2j+1 = N$, under the  $SU(2)$ group of rotations. 
      
     We compute the total number of modes for each generation in  figure \ref{LeftAndRight}.
  
    \vspace{ .2 cm}
\begin{figure}[h]
  \begin{center}
  	 \begin{tabular}{||l | c |c ||}\hline \hline
  {\rm Field} & $SU(3) \times SU(2) \times U(1)$ & {\rm Number of 2d modes (left - right)} \\ \hline 
  $q_L$ & $({\bf  3}, {\bf 2})_{{1\over 6}~~}$ & ~~~~Q \\
  $u_R$ & $({\bf 3}, {\bf 1})_{{2 \over 3}~~} $ &  - 2 Q \\
  $d_R$ & $({\bf  3}, {\bf 1})_{-{ 1 \over 3} } $ & ~~~~ Q \\
  $ l_L $ & $({\bf  1}, {\bf 2})_{- \half} $ & ~~~- Q \\
  $ e_R $ & $({\bf  1}, {\bf 1})_{-1} $ &~~~~ Q \\ \hline \hline
  	 \end{tabular}
  	\caption{ Total number of complex two dimensional modes from each generation. For example, for the left handed quark doublets we have $Q/6$ two dimensional leftmoving fermions in the representation $({\bf 3},{\bf 2})$ giving a total number of $Q$ complex fermion fields.  }
  	\label{LeftAndRight}
  	  \end{center}
\end{figure}
We see that the total number of fields per generation is $3Q$ left and right moving fields. We then have  $9Q$ fields for the three generations. Notice that we get the same total number of left and right moving fields, as required by the cancellation of the 2d gravitational anomaly. Of course, all two dimensional anomalies vanish since  the four dimensional anomalies vanish.    
 
 In summary, inside the electroweak corona the fermions lead to  the two dimensional massless   fields in table 
 \ref{LeftAndRight}. They move along the radial direction. All other modes, coming from higher Landau levels,  have higher energies of order $|F|$. In this region, the Higgs field also has energies of order $|F|$. 
     
     The dynamical gauge fields can still lead to long distance effects. 
     They propagate along all four dimensions but they are interacting with  fermions that move along two dimensions. These can lead to interesting effects. 
     
     For example, if we just had a $U(1)$ gauge field, then we can analyze the problem as follows \cite{Thompson:1998ss,Gralla:2018bvg}. 
     We bosonize the two dimensional fermions and there will be one overall boson mode that interacts with the gauge field. The combined dynamics of the $U(1)$ gauge field together with this boson leads to a long distance theory called ``force free electrodynamics'' (see \cite{Gralla:2014yja} for a review), which can be described in terms of the gauge field. In the process we also produce excitations 
     which have energies $m^2 \sim g^2 |F|$ coming from a version of the Higgs mechanism. A more detailed study  in \cite{Gusynin:1999pq}  argues that  also a much smaller non-perturbative mass is generated.  The fact that we can use this force free electrodynamic description implies that a rotating black hole would lose energy relatively rapidly  \cite{Blandford:1977ds} (see \cite{Gralla:2014yja} for a review). The power radiated scales as 
     $P \propto Q^2 \Omega^2$ \cite{Gralla:2014yja}. For a rapidly rotating black hole, $\Omega \sim 1/r_e$,  we find a decay timescale $\tau =  M/P \propto Q l_p$. 
          
     Since the region with restored gauge symmetry can be large, 
     $r_w, r_h \gg 1/m_H$, we should also analyze the IR dynamics of the    $SU(2)$ and $SU(3)$ gauge fields. The complete analysis of the infrared behavior of the theory is beyond the scope of this paper. See 
     \cite{Kharzeev:2012ph,Chernodub:2012tf,Shovkovy:2012zn} for reviews on QCD in high magnetic fields\footnote{ For $Q=1$ GUT monopoles the effects of QCD  were discussed in \cite{Craigie:1983vq}.}. 
   
   As we move outside the corona, some fields are still charged under the electromagnetic $U(1)$ gauge field, such as the electron. These fields are able to traverse the corona with no impediment \cite{Ambjorn:1989sz}. Outside the corona the electron  has the usual electron mass.
         However, since the electron mass is small,  the effects of the magnetic field dominate up to the distance where $|F|\sim m_e^2$.
          Other fields, such as the top quark, will get a large mass in the outside region but they continue to have the factor of $Q$ in their degeneracy. For neutral fields, such as the neutrino, we lose their large degeneracy outside the corona, but they can still exist as light four dimensional fields outside.  
  
  The large magnetic field also has important effects for the strong interactions. In fact,  up to distances where   $|F| \sim \Lambda_{QCD}^2 $,   the effects of the magnetic field are very important and distort the QCD vacuum.  Aspects of QCD for large magnetic fields were reviewed in e.g. \cite{Kharzeev:2012ph,Chernodub:2012tf,Shovkovy:2012zn}.  So, the black hole will also have a ``QCD corona'' where the vacuum has large deviations away from the usual four dimensional confining vacuum.       
   
 \section{Evaporation of near extremal black holes  } 

So far,  we have discussed  extremal, zero temperature black holes. 
Now we will increase their mass above extremality, which will raise their temperatures and we will explore the effects of Hawking radiation. 

Now, there are two important parameters. The first is the charge $Q$, which will determine all the extremal properties are we discussed above. Then we have the deviation away from extremality, which we  parametrize in terms of the black hole temperature, related to the mass above extremality by 
\be \la{MassAbEx}
 M- M_e \sim 2 \pi^2  { r_e^3 T^2 \over l_p^2 }  = { 2 \pi^{7/2} \over {g'}^3 }  Q^3  T^2 l_p 
\ee
This formula holds for temperatures lower than about $T < 1/r_e$. 

Since the black hole has a somewhat intricate structure, we will just illustrate some of the more salient phenomena.  

In the region where the electroweak symmetry is unbroken we 
can use a two dimensional formula to compute the flux of energy out of the black hole
 \be \la{BHPower}
P= { d E \over d t } = { c \pi \over 12 } T^2 = { 3 Q \pi \over 4 } T^2 ~,~~~~~~~c = 9 Q
\ee 
where we used the number of modes from table \nref{LeftAndRight}. This result is correct for temperatures high enough that we can neglect the effects of the dynamical $SU(2)$ or $SU(3)$ gauge fields. 

If this energy were to make it all the way to infinity, then 
we would be able to combine it with \nref{MassAbEx} to find an exponential decay in the mass of the form 
$M - M_{e} \propto e^{ -t/\tau} $ with 
\be \la{DecayTime}
\tau   \sim { 8 \pi^{5/2} Q^2 l_p \over 3 {g'}^3}  
 \ee 
Compared with the evaporation timescale of a Schwarzschild black hole of the same radius, it is smaller by a factor of $1/Q$. 

For example, we can consider a magnetic black hole with mass 
of order $10^{12} $ Kg which, for a Schwarzschild black hole, would decay within the age  of the universe. Here we would get $\tau \sim $ of order a couple of milliseconds (this mass   corresponds to $Q \sim 10^{18}$). For an order one deviation from extremality, i.e.  ${ M - M_{ext} \over M} \sim o(1) $, this would have a temperature of order a few GeV.  

These decay rates however, are not the whole story since for  
 charges $Q< Q_{ew}$ we need to include effects of the electroweak corona. Depending on the temperature, this can act as a reflecting mirror for some modes since the fields will have a non-zero mass outside. Also, for the neutrino, the magnetic field does not give rise to $Q$ modes once we are outside the corona radius. These effects would reflect some of the energy back into the black hole. Therefore the true evaporation rate will depend on the amount of energy that can make it past the electroweak corona.
 We expect that for temperature larger than the mass of the electron, $T> m_e$, the energy flux carried by the electrons will make to infinity, or to a large enough radius, where positrons and electrons annihilate, etc. If we assume that only electrons make it out, then the power is \nref{BHPower} with $c =Q$. 

If the black hole mass or charge is larger than \nref{EWtrans}, then there will be no electroweak corona. For near extremal black holes the temperatures will be always lower than about $1/r_e$. For these cases the emission is very slow, not enhanced by a factor of $Q$. If we consider black holes with an order one deviation from extremality, or 
 $T \sim 1/r_e$, then we expect a big jump in the energy radiated when the temperature becomes of the order of the electron mass. It is here that that the additional factor of $Q$ kicks in. This is a charge of about 
 $Q \sim 10^{21}$, and mass of order $10^{14}$ kg. For these relatively high temperatures, we expect the estimates in \nref{BHPower} \nref{DecayTime} to be correct up to  a correction due to the number of species that can make it out to infinity.  

  For temperatures much less than the electron mass the large rate we found above would be multiplied by an extra factor $e^{-m/T}$.  In addition, at these low temperatures, it would also be important to understand the precise nature of the vacuum inside the corona,  after taking into account the gauge interactions.  Assuming that these do not generate a gap for all fields inside the corona,  we can estimate the emission rate as follows.   We expect that the whole region inside the corona will heat up to the temperature $T$ of the black hole. Then we could imagine computing emission from the corona as a black body at this temperature. If we use the four dimensional black body formula we would get an energy loss of the order 
 \be  \la{LB}
 { d E \over d t } \sim   4\pi r_w^2 { 2 \pi^2 \over 90}  T^4 
 \ee 
 where $r_w$ is the outer corona radius \nref{rcdef}. This includes only radiation into photons, but we should also want to include neutrinos.  We will not analyze this in detail, since  \nref{LB}  is just a very crude estimate of a lower bound for the energy emitted.

 \subsection{ Energy production } 

If these black holes existed, and we could capture them, we could imagine using them to catalyze proton decay, in the same way that we could use unit charge magnetic monopoles.\footnote{
To prevent them from falling in the gravitational field of the earth, we would need a magnetic field of about a Tesla. } 
For relatively small black holes, the evaporation rate is very fast, so that the limiting factor seems to be the rate at which we can get matter to fall into the black hole. 

As a fun example, consider a black hole of a mass of 100 kg, the charge is about $Q =10^9$. If the temperature is above the electron mass, then 
\nref{BHPower}, with $c=1$,  would give us $10^{16}$ Watts. This looks like a huge power, but it comes out in a time $\tau \sim 10^{-21} s$, during which the total energy that comes out is about 10 microjoules.
However, as the temperature drops below the electron mass the energy flux  reduces exponentially. So that we can imagine an equilibrium between the ingoing baryonic matter and the outgoing energy at a rate that is basically set by the rate of ingoing matter. For a black hole of this mass, the QCD radius is about the size of an atom. We expect that once we get the baryons to this radius, they will get into the black hole without much difficulty. So if one manages to get $10^{16}$ protons a second one could get about a megawatt.  Of course, we just superficially sketched some estimates here.
  
As a comparison, we could also imagine using Hawking radiation from a  Schwarzschild (uncharged) black hole. The power is also proportional to the square of the temperature $P \sim T^2$. But,  in contrast to \nref{BHPower}, without the factor of $Q$. In addition, the specific heat is negative. For example, we get a megawatt with a black hole whose mass is of order $10^{13}$ kg, at a temperature of the order of the electron mass.   A 100 kg black hole would evaporate in about $10^{-12}$ seconds.

\section{ Production in the early universe? }

Of course, the fact that these are solutions of the Standard Model does not mean that they are easy to produce. In fact, they seem harder to produce than individual  magnetic monopoles. 

Producing these black holes artificially looks very difficult, since it would involve producing or gathering lots of magnetic monopoles and then collapsing them into a black hole, even though same sign monopoles repel each other. 

Could they have been produced in the early universe?
It is possible to produce uncharged black holes.  For example, one could have larger primordial fluctuations at some specific length scales that, after inflation,  produces primordial black holes, of any desired size
\cite{GarciaBellido:1996qt} (see \cite{Garcia-Bellido:2017fdg} for a review). 

Producing charged black holes seems harder. Producing them during inflation is unlikely \cite{Bousso:1996au}.   One plausible mechanism (similar to the one discussed in \cite{Bai:2019zcd,Stojkovic:2004hz}),  would be to produce first a large number of monopoles and anti monopoles. Then at larger scales we have large primordial fluctuations that produce black holes. If the black hole swallows $N$ monopoles or anti-monopoles, then one might expect a  net charge of order $\sqrt{N}$. If their masses are small enough, they would evaporate quickly to extremality. These extremal black holes can survive till today. Of course, we would need to make sure that there are not enough left over monopoles to cause trouble.

 
 One could also wonder whether they could be the dark matter, since they cannot decay. This issue was studied for magnetic monopoles, and it was found that the most stringent bound comes from baryon decay catalysis in astronomical bodies, e.g. neutron stars, see \cite{Turner:1985be} for a review. It would be interesting to see whether this is changed if we assumed that they are magnetic black holes, which would make collisions with astronomical bodies rearer, but with possibly more spectacular results once the collisions happen. Another constraint comes from the Parker bound on neutralizing the magnetic field of the galaxy \cite{Turner:1982ag}. We will not attempt to derive concrete bounds here. 
  
  Primordial extremal black holes in the dark sector were proposed as dark matter  candidates  in \cite{Bai:2019zcd}. 
 
  \section{Discussion} 
  
  Here we have pointed out a few  peculiar features of magnetically charged black holes. 
  
  \begin{itemize} 
  \item They can be very long lived, even with relatively low masses. For example, they can last for the age of the universe for masses larger than about 0.1 kg if unit charge magnetic monopoles have the mass they have in a GUT theory.
  \item They can have very large magnetic fields that restore the electroweak symmetry  around the black hole. This region can be relatively large. This could enable us to explore features of the Higgs potential that are hard to access at colliders.   Similarly there is a region where the QCD vacuum is distorted. 
  \item Hawking radiation effects are enhanced by a factor of $Q$, leading to relatively rapid decay.	
  \item These black holes would offer a window to very high energy physics, due to their high magnetic fields. 
  \end{itemize}

There are a number of questions we have not discussed adequately.  

\begin{itemize}
	\item Can they be created in the early universe, via a plausible mechanism?
	\item Could they be the dark matter? Or a fraction of the dark matter?
	\item What are their astrophysical signatures?
	 
\end{itemize}

Some of their properties are common with magnetic monopoles. In some sense, we can view them as very high charge bound states of monopoles, so that their astrophysical constraints would be similar in spirit to those of monopoles, see \cite{Bai:2020spd} for a recent dicussion. 

Due the presence of light fermions moving radially along magnetic field lines, these objects are vaguely similar to black holes pierced by superconducting strings. Here the ``strings'' are just the magnetic field lines.

   

\subsection*{Acknowledgments}

We would like to thank  Nima Arkani-Hamed, Asimina Arvanitaki, Adam Brown, Savas Dimopoulos,  Isabel Garcia-Garcia,  Sa\v{s}o Grozdanov, Ted Jacobson, Alexander Milekhin, Fedor Popov, G. Santos, Nikolay Sukhov, Ken Van Tilburg,   Eric Weinberg, Edward Witten and Matias Zaldarriaga for discussions.   

J.M. is supported in part by U.S. Department of Energy grant DE-SC0009988 and by the Simons Foundation grant 385600. This work was partly done at the KITP during  programs on  ``Gravitational Holography" and ``From Inflation to the Hot Big Bang" and  was supported in part by the National Science Foundation under Grant No. NSF PHY-1748958.

\appendix

\section{Qualitative aspects of the electroweak corona }

 \begin{figure}[h]
    \begin{center}
   \includegraphics[scale=.29]{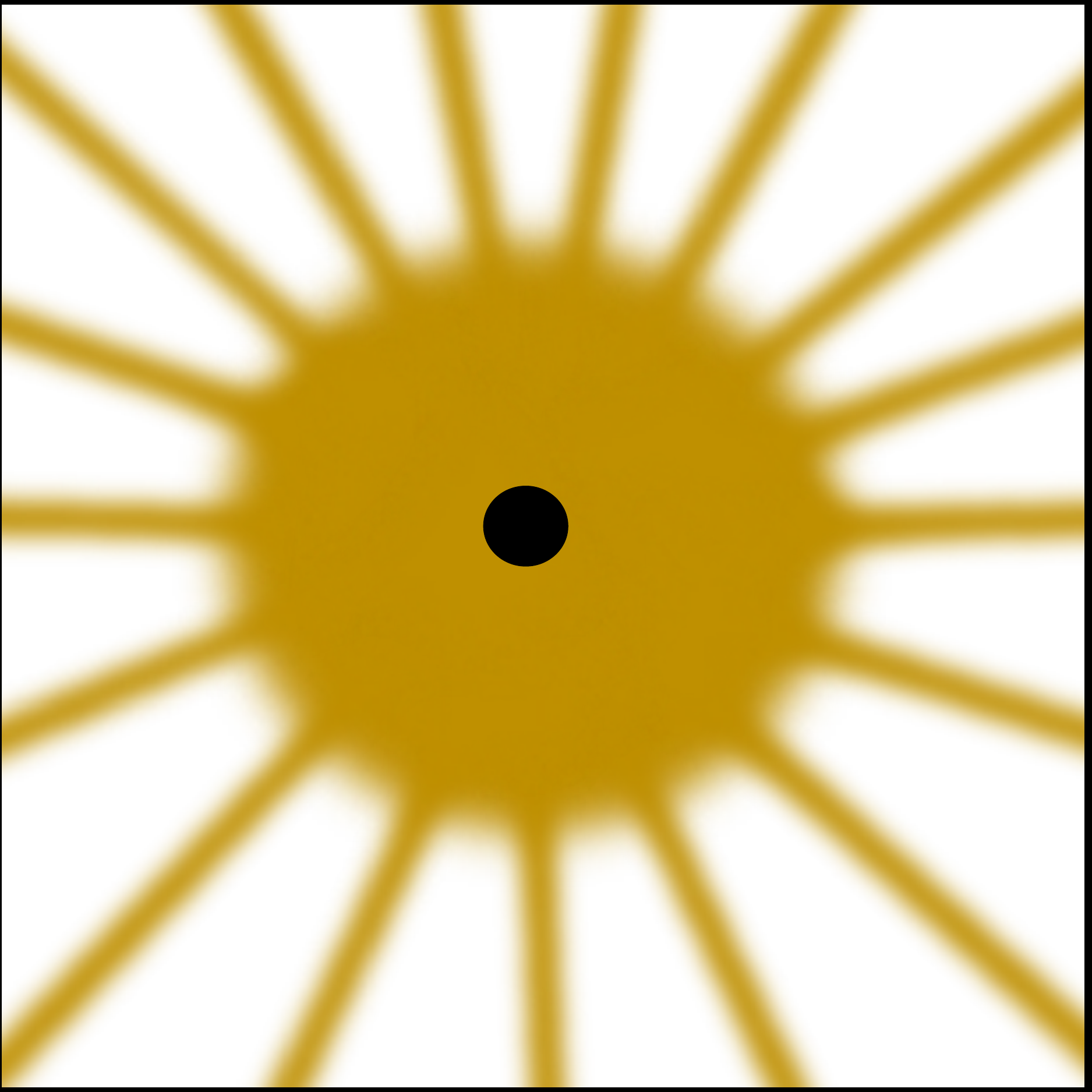}  ~~~~~~~~~~~~ 
    \includegraphics[scale=.3]{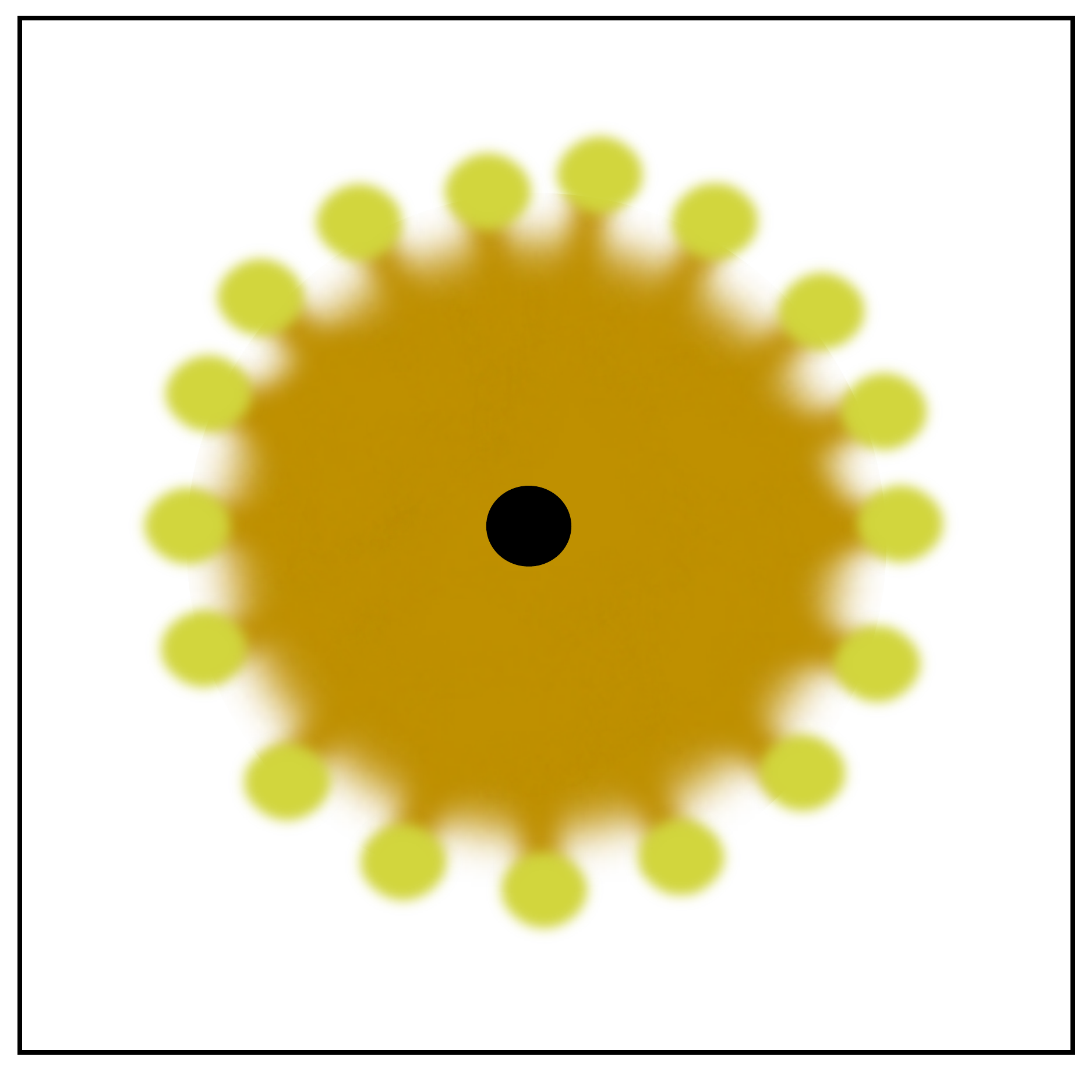} \\
  ~~ (a) ~~~~~~~~~~~~~~~~~~~~~~~~~~~~~~~~~~~~~~~~~~~~~~~~~~~(b)
    \end{center}
    \caption{ (a) Hairy magnetic black hole in the abelian Higgs model with a central region where the Higgs field is small and an outside region where it forms  vortex strings.   (b)  Once we add the $SU(2)$ gauge fields, the  vortex strings can end on monopoles.       }
    \label{BHHiggs}
\end{figure}

 As a first toy problem one could consider a magnetic black hole in an abelian Higgs model. Namely a model with  $U(1)$ gauge field and a charged scalar field with a Mexican hat potential, whose scale is much less than $M_p$. In this case, for some range of charges, the symmetry is restored in the near horizon region and as we go away, the magnetic field becomes confined to strings or vortices, which would repel each other or attract each other depending on the ratio of the quartic coupling to the electric charge. For the case when they repel, the solution would take a form as in figure \ref{BHHiggs}(a). 
  This toy model is what we would get if we removed the $SU(2)$ gauge fields from the usual electroweak theory\footnote{The strings in figure \ref{BHHiggs} would have two units of hypercharge flux, in our normalization, since the Higgs field has hypercharge $1/2$.}.

 If we now add back the $SU(2)$ gauge fields, then these strings can end on $SU(2)$ monopoles that create then an ordinary magnetic field. And we  have configurations as in figure \ref{BHHiggs}(b). These are just qualitative figures emphasizing their topology.   
 It seems possible to have a variety of configurations depending on the parameters of the model.  The strings ending on monopoles   are reminiscent of the $Z$ strings discussed in 
 \cite{Nambu:1977ag}, see \cite{Achucarro:1999it} for a review.   They are not identical because the hypercharge flux on each string is two here, and the monopole charge is also two, as in the solutions in \cite{Cho:1996qd}.  
 
  For the physical values of the parameters we expect that the corona has the following form. At each radial position we expect to have a configuration similar to the electroweak configuration with a constant magnetic field discussed in \cite{Ambjorn:1989sz}, which consists of a lattice of vortices. As we move in the radial direction the average magnetic fluxes decreases and we interpolate continuously between the $h=0$ vacuum and the $h\not=0$ vacuum.

\bibliographystyle{apsrev4-1long}
\bibliography{RevisedSubmitted.bib}
\end{document}